# Primitive Virtual Negative Charge

*Kiyoung Kim* [1]

**Abstract**

Physical fields, such as gravity and electromagnetic field, are interpreted as results from rearrangement of vacuum particles to get the equilibrium of net charge density and net mass density in 4-dimensional complex space. Then, both fields should interact to each other in that physical interaction is considered as a field-to-field interaction. Hence, Mass-Charge interaction is introduced with primitive-virtual negative charge defined for the mass. With the concept of Mass-Charge interaction electric equilibrium of the earth is discussed, especially about the electric field and magnetic field of the earth. For unsettled phenomena related with the earth's gravity, such as antigravity phenomenon, gravity anomalies during the solar eclipses, the connection between geomagnetic storms and earthquakes, etc., possible explanations are discussed.

## 1. Introduction

In physics, the connection from classical theories to quantum theory and theories of relativity is not natural but enforced in the name of facts or phenomena. To find a logical connection a unique fundamental theory was searched with ontological inquiries of physical quantities and physical laws.

With Occam's razor in mind, the fundamental theory [1] was suggested to understand the physical theories comprehensively. As a model of physical space, 4-dimentional complex space was suggested, in which the Imaginary Space of the 4-D complex space is filled with vacuum particles. The vacuum particles are confined with a negative energy as like $\varepsilon_v \approx -m_e c^2$ and doing harmonic interactions among them. Physical objects keep interacting with those vacuum particles, and physical effects are realized through $U(1)$ symmetry in phenomena. The interactions with vacuum particles result in physical fields in gravitation and/or electromagnetism. For a dynamic state of physical object, the interactions with vacuum particles are following the physical object in a form of wave packet in the Imaginary Space. Here, the concept of wave packet is similar to the pilot wave in pilot wave theory or hidden variable theory [2].

---

[1] kkim.pvn@gmail.com

Once the intangible reality – vacuum particles -- is introduced in physical vacuum, both theories in physics -- quantum physics and relativity -- can be understood comprehensively as following: The wave function in quantum physics is the realization of physical effects of vacuum particles, which is the intrinsic property of statistical nature. Length contraction, time dilation, and mass increase in special theory of relativity are not for the physical object itself but for the physical effects came from the interactions with vacuum particles. With this model of physical space, the Schrödinger equation in quantum physics was derived [3].

Without loss of generality in the model of the physical space, the charge of the vacuum particles in the Imaginary Space can be changed from negative charge to positive charge. Hereafter, the vacuum particles are supposed to have negative mass, such as $\varepsilon_v \approx -m_e c^2$, positive charges and spins, such as in positron $e^+$. The reason for the change of charge sign for vacuum particles is to adapt to the macroscopic phenomenon on the earth, which is the direction of electric field on the earth's surface.

As an ontological entity underlying physical phenomena, physical fields -- gravitational field and electromagnetic field – have a unique mechanism, in which the physical fields are come from the rearrangement of vacuum particles through the interactions with the physical object that has mass or charges. Here, the rearrangement of vacuum particles should be made to minimize the changes from the equilibrium of net-mass[2] density and net-charge[3] density in the physical space.

Physical object – with charges or mass – interacts with those vacuum particles in the Imaginary Space, and the results of interactions are appeared as physical fields in phenomena. For instance, if a massive object is introduced in the space, the massive object attracts vacuum particles to minimize the change of net-mass density in the space. It is the same principle for charged particles as for the massive object. Positive charge repulses vacuum particles and negative charge attracts vacuum particles to minimize the change of net-charge density in the space. On the other hand, magnetic field is come from the aligned spins of vacuum particles to minimize the change of net current density for a dynamic state of physical charges in the space. In which the net current density is considered with the spins of vacuum particles and physical current, and the direction of spin alignments should be normal to the physical current. Here, the vacuum particle is supposed to have a finite size with spinning. Lorentz force also can be explained with the same reasoning as before.

Considering that gravitational field and electric field, both have same ontological entities in 4-D complex space and that physical interactions are results from field-to-field interactions, Mass-charge interaction should be existent intrinsically in nature.

In phenomena, gravitational interaction is much weaker than electric interaction because when a massive object is introduced in the space, the rearrangement of vacuum particles is initiated to get a net-mass density equilibrium in the space, but the charges of vacuum particles react against the rearrangement. Since the order of magnitude of vacuum particle's charge is much bigger than the order of magnitude of its mass[4], the effects from the vacuum particle's charges are dominant in the process of the rearrangement. With this fact, it can be explained why gravitational interaction is attractive. In the gravitational interaction between two massive objects, vacuum particle's charge distribution is getting more stable when the two massive objects are close than apart to each other.

---

[2] net-mass means physical mass and background mass in the imaginary space.
[3] net-charges means physical charges and background charges in the imaginary space.

For a massive object, the equivalent charge on mass-charge interaction can be defined for the gravitational field produced by the mass. Let's name it as primitive-virtual negative (PVN) charge. To estimate the PVN charge, firstly, the interaction strengths of electrostatic interaction and gravitational interaction should be compared. In 4-D complex space, the rearrangement process of charges and masses of vacuum particles should be compared because a field of physical charge or a field of physical mass is generated by the vacuum particles in the Imaginary Space. However, up to now there is no absolute criterion to compare the strengths of physical interactions in physics although the strength ratio $R$ of electromagnetic interaction to gravitational interaction has been known as $R \approx 10^{36-44}$ [4].

For a rough estimation, there is another way to estimate the PVN charge on a massive object. As a criterion to compare the interaction strengths, proton (nucleon) can be chosen as the basic units of mass and charge as following: In nuclear physics; the mass of an object is determined by the number of nucleus in the object, and the mass of nucleus is determined by the number of nucleons in the nucleus, in which average binding energy per nucleon is less than 10 MeV [5]. Nucleons are proton and neutron, in which proton ($m_p \approx 938$ MeV) is almost identical to neutron ($m_n \approx 939$ MeV) except the electric charge ($q_p \approx |q_e|$) [6]. Therefore, the mass of a massive object should be proportional to the number of nucleons in the object in that the average binding energy per nucleon is 2 orders smaller than proton's mass energy and the mass difference of proton and neutron is 3 orders smaller than the proton's mass.

The strength ratio of gravitational interaction to electrostatic interaction for proton-proton is $8.1 \times 10^{-37}$. The ratio of field strengths ($R$) made by each proton and its charge can be interpreted as $R \sim 10^{-20}$. Hence, the gravitational field strength by mass $M$ is compared to the electrostatic field strength by charge $Q \sim -10^{-20} M$ in its magnitude ($G \sim 10^{-20} k_c$ in MKSA). Since the PVN charge is not a physical charge, the coupling constant in mass-charge interaction should be compared with the coupling constant in electrostatic interaction.

Considering that physical field is the result of rearrangement of vacuum particles to get the equilibrium in net-mass density and net-charge density in vacuum, the field strength should be determined by the charge-to-mass ratio of vacuum particles and sources, such as physical mass or physical charge. If the coupling constants in gravitational interaction and electrostatic interaction is expressed using the charge-to-mass ratio $R_{em}$ of vacuum particles, which is given as $R_{em} \equiv \left| \frac{e}{m_e} \right| = 1.76 \times 10^{11} \left( \frac{C}{kg} \right)$, gravitational constant $G = 1.17 \times 10^1 \times R_{em}^{-1}$ and electrostatic constant $k_c = 5.13 \times 10^{-2} \times R_{em}$. Here, the constants, $1.17 \times 10^1$ in G and $5.13 \times 10^{-2}$ in $k_c$ can be interpreted as the source contributions. Therefore, the coupling constant in mass-charge interaction can be estimated as $C_{mc} \sim 1.5 \times 10^{+1} \times k_c$. Including this mass-charge coupling constant that is about 1 order higher than electrostatic constant, the PVN charge of mass $M$ can be estimated as

$$Q_v[\text{C}] \sim C_{pvn} \cdot M[\text{kg}] \tag{1}$$

at least for macroscopic phenomena. Here, the conversion constant $C_{pvn} \equiv 10^{-19} [\text{C}/\text{kg}]$, in which the ratio of field strengths was used in the proton-proton interaction. The distribution of vacuum particles for a massive object is similar to the distribution for a negative charge. Although there is no physical charge in the gravitational field, the field interacts with electric charges because physical interaction is field-to-field interaction.

---
[4] It should be bare charge and bare mass for vacuum particles; however, the charge and mass of positron can be used for a rough estimation.

However, it is not easy to estimate the primitive-virtual mass for an electric charge because the effect of charge of vacuum particles is much bigger than the effect of mass in the rearrangement process of vacuum particles, and thus the physical electric charge cannot be ignored in the rearrangement process of vacuum particles. For the same reason, gravitational interaction cannot be substituted to electric interaction, and vice versa.

Even though many fields in physics should be reviewed with this new interpretation of physical interactions, first, let us interpret physical interactions in the Imaginary Space.

## 2. **Physical Interactions in Imaginary Space**

First of all, we can interpret what is the inertial force in Newtonian mechanics, which is known as fictitious force. When a physical object, such as rigid body, is moving with velocity $\upsilon$, the physical momentum for the moving object is expressed as $p = m\upsilon$. All of sudden, if an external force is applied to the object, there is the inertial force exerted on the object. The magnitude of inertial force is exactly same as the magnitude of external force, but in opposite direction to the external force.

Vacuum particles are interacting with the physical object; if the physical object is in motion with velocity $\upsilon$, the interaction is following the physical object in a form of wave packet with momentum $p = m\upsilon$ [1]. If an external force is applied on the object, the wave packet reacts against the change enforced by the external force. The reaction of the wave packet appears as the inertial force on the physical object.

Physical forces, such as electric force, magnetic force, and gravitational force, in phenomena are results from the enforcements on the physical object to change the position in space and time by the physical fields, in which the physical fields are being changed through field-to-field interactions. By the way, strong interaction and weak interaction, which are effective only in short ranges, are considered as special cases of electromagnetic and gravitational interactions.

As mentioned above, gravitational field and electric field can interact to each other in the physical space. To know if magnetic fields can also interact with gravitational fields or not, we need to know the relations between the spin alignments of vacuum particles and the corresponding magnetic field since magnetic field is related with spins of vacuum particles

In Maxwell's equations, the equations still hold when the electric fields and the magnetic fields are exchanged as $\vec{E}' = K\vec{B}$ and $\vec{H}' = -K\vec{D}$ ($K$ is a constant) in free space [7]. The duality of Maxwell's equations in phenomena consists with the generation process of physical fields in the 4-D complex space. Furthermore, it is clear that there is no magnetic monopole in general.

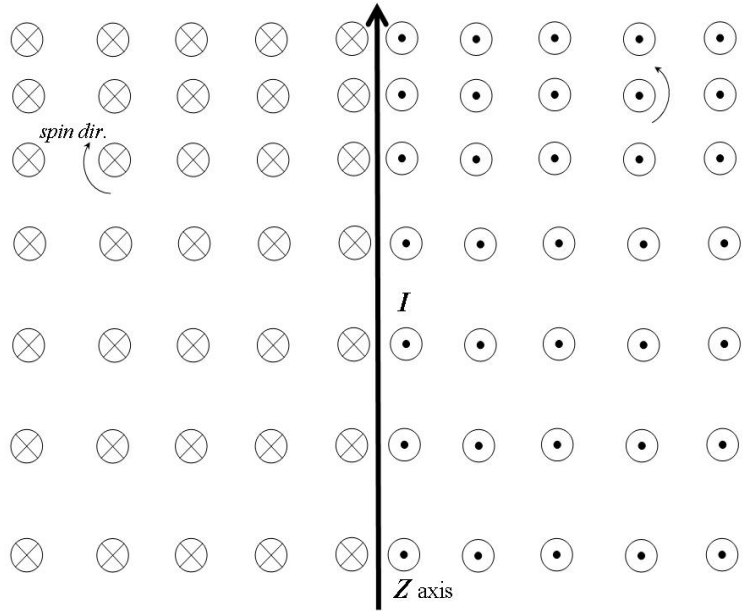

Figure 1. Magnetic field around a current wire and rearrangement of vacuum particles.

With a simple case, let us interpret how vacuum particles are rearranged in the physical space against a physical current. Fig. (1) indicates the rearrangement of vacuum particles near the physical current in a wire. In phenomena, electric fields are only in Z-direction and magnetic fields are only in $\phi$-direction in cylindrical coordinates. It is noticeable that the spin alignment of vacuum particles is in the opposite direction from the magnetic field direction in phenomena.

However, the vacuum particles are pushed toward the electric fields; the spins of vacuum particles are aligned to minimize the effect of the physical current in physical space. The electric field $E$ is proportional to the tension of vacuum-particle-strings in Z-direction, and magnetic induction $B_\phi$ is proportional to the spin number density in a vacuum-particle-string that is circular closed and consists of spins aligned in $\phi$-direction.

If $\rho$ is large enough as $\rho_1 > \rho_2 \gg 0$, $B$ or the spin number density in $\phi$-direction is inversely proportional to the radial distance $\rho$ and directly proportional to the current $I$ because $E_\rho = 0$ and the size of vacuum particle can be ignored from a macroscopic point of view. Thus, magnetic induction $B_\phi \propto \dfrac{I}{2\pi\rho}$.

Now, if there is a background magnetic induction $B_b$ and the direction is out of paper, the spin alignments of vacuum particles are the same as the alignments in left hand side (LHS) from the current $I$ in the Fig. (1). The number of spin density in the LHS is higher than in right hand side (RHS). Remembering that the rearrangement of vacuum particles is the result from the reaction against any change from the equilibrium in physical space, there should be a force exerted on the current wire to the RHS. The magnitude of force is proportional to the difference of spin number density in both sides, in which the difference is directly proportional to $B_b$ and $I$. In electromagnetism it is known as magnetic force, $\vec{F}_m = (\vec{I} \times \vec{B}_b)l$ on current wire with length $l$. As above, the magnitude of magnetic induction is proportional to the spin number density in vacuum-particle-strings, but it is not related to the tension of the strings. Because magnetic induction $B$ is appeared as the spin alignment of vacuum particles responding to physical currents, the direction of the magnetic induction $B$ is always normal to the current direction or the electric field generating the current. If there is no physical current but a local electric field being changed with time as $\dfrac{\partial \vec{E}}{\partial t} \neq 0$, the rearrangement of vacuum particles is against the change of electric field. Thus, the number density of vacuum particles is increasing in the direction of the electric field. This rearrangement of vacuum particles is equal to a local current in the physical space, and thus surrounding vacuum particles are also responding to the change as aligning their spins. Hence, $\nabla \times \vec{B}$ in Maxwell's equations has the term of vacuum displacement current as $\vec{D} \propto \dfrac{\partial \vec{E}}{\partial t}$.

If a negative charge $-Q$ [5] is moving under the influence of uniform $\vec{B}$ field, Lorentz force is exerted on the charge as $\vec{F} = (-Q)(\vec{v} \times \vec{B})$ in MKSA. Fig. (2) shows how vacuum particles respond on the moving charge $-Q$ with the $\vec{B}$ field. The direction of magnetic induction $\vec{B}$ is out of paper since the reaction of vacuum particles is against the change in physical space. Fig. (2) shows that some vacuum particles are attracted to the negative charge, and it makes a pilot wave moving with velocity $\vec{v}$. With the charge moving, the electric field in the space keeps changing. Vacuum particles also react against the change of electric field to minimize the current effect of the moving charge. In Fig. (2), the spin number densities of vacuum particles surrounding the charge are different in LHS and RHS from the moving charge. The reaction of vacuum particles for this number density difference appears as a force exerted on the charge, in which the force is the Lorentz force in phenomena. The difference of the spin number density is directly proportional to the velocity $\vec{v}$ and background spin number density, which is also directly proportional to the intensity of $\vec{B}$.

---

[5] For easy drawing negative charge was chosen, and $Q > 0$.

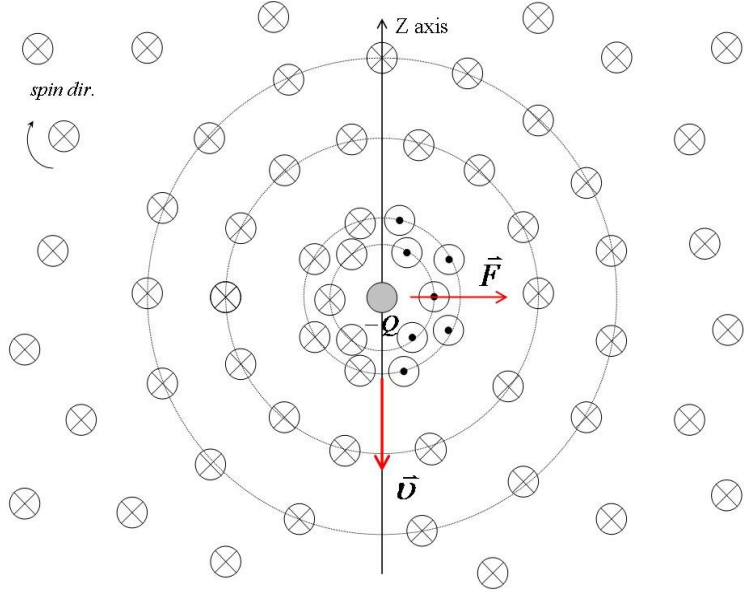

Figure 2. Lorentz force on a moving charge.

There are two possible ways to propagate energy in physical vacuum, which is the Imaginary Subspace in 4-D complex space model. One is transverse mode and the other is longitudinal mode of vacuum-particle-string. The relation of $\vec{E}$ and $\vec{B}$ in the transverse mode, which is the way of light propagation [1], can be understood with the spin reactions of vacuum particles to the electric fields generated by the sinusoidal motion of vacuum-particle-strings. In the longitudinal mode; the disturbance of electric fields is in vacuum-particle-strings and oscillating in the longitudinal direction. The spins of surrounding vacuum particles react for the disturbance and generate magnetic fields around the vacuum-particle-strings, in which the magnetic fields are also oscillating with the electric fields. For instance, if the electric field is increasing, the magnetic field around the string is in right handed direction to the direction of propagation; if the electric field is decreasing, the magnetic field is in left handed direction. If a pulse is considered in the longitudinal propagation, a magnetic pulse is also accompanied. If the pulse is created with other particle ($e^{\pm}, \mu^{\pm}$, etc.) that has angular momentum in the longitudinal direction, the pulse should have helicity because of energy-momentum conservation.

If there is no disturbance in physical space with physical charge or mass, electric fields and gravitational fields cannot be distinguished. Once a disturbance is given to the physical space, basically both modes of propagation, transverse mode and longitudinal mode, should be possible as long as there is no constraint.

Electromagnetic interaction should be in transverse mode since the time dependent electric fields in the space always accompany with the change of magnetic fields, and vice versa. Gravitational interaction should be in longitudinal mode because magnetic fields are not related in direct as in the electromagnetic interactions. However, the change of gravitational field creates a change of electric field and then the change of electric field can also produce a magnetic field.

Weak interaction, which is one of fundamental interactions in nature, should be in longitudinal mode because it has been known that electromagnetic interaction is highly suppressed in the interaction range ($\sim 10^{-18}$ m) [4]. If weak interaction is in longitudinal mode, it is natural to suppose that neutrino, one of elementary particles, is a longitudinal pulse because it appears only in weak interaction.

## 3. **Mass-Charge Interaction**

Following example shows that mass object can interact with charges. In Fig. (3), there is a spherical shell with radius $R$ that is nonconductive and has no mass. In the beginning, let us say, there was nothing (free space) inside the spherical shell except an equal number of positive and negative charges just below the surface of the spherical shell at radius $R$. Now, a nonconductive physical object is put at the center, in which the object has mass $M$ and radius $r$ in the shape of sphere.

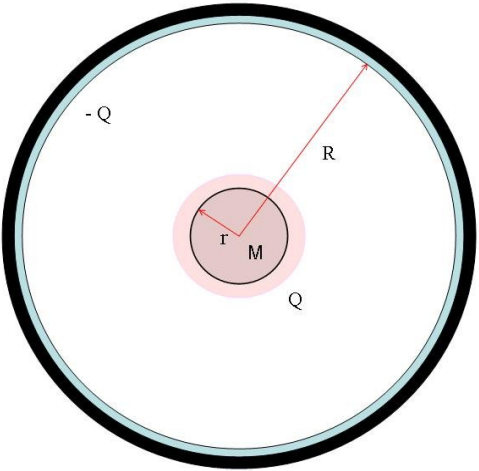

Figure 3. Mass-charge interaction in Gravity.

For the mass object, the PVN charge $Q_v$ is given as $Q_v = C_{pvn} M$. Hence, positive charges are pulled down to the center and distributed uniformly at around surface of the mass object until total amount of positive charges $Q$ is equal to the $|Q_v|$. Then, the electric field becomes to zero in the space between radius $r$ and radius $R$ since the net charge inside the spherical shell (radius $< R$) is zero. The electric field energy density in the space is also zero; hence, gravitational force is also disappeared. In this process, the work on the space done by the PVN charge $Q_\upsilon$ is positive and equal to the change in the electric potential energy of charge $Q$ in absolute value.

$$\Delta U_{QQ_\upsilon} = -\frac{Q^2}{8\pi\varepsilon_0}\left(\frac{1}{r}-\frac{1}{R}\right) \qquad (2)$$

In other words, the tensions among vacuum particles corresponding to the space has been released as in free space. If the charge $Q < |Q_\upsilon|$, then $|Q'_\upsilon| = |Q_\upsilon| - Q$ and the effective mass for gravitational force $M' \propto 10^{19}|Q'_\upsilon|$. The relation of the effective mass to the electric field inside the spherical shell is

$$\frac{M'}{M} = \frac{Q'_\upsilon}{Q_\upsilon} = \frac{E'}{E} \qquad (3)$$

## 3.1 Mass and Charges (stationary state)

Once the PVN charge is introduced for mass, the evaluation of electric field is direct because there are only two kinds of charges, such as positive charges, and negative charges including the PVN charges. The linearity, which means the superposition principle in physics, and locality still can be assumed to be valid. On the other hand, the evaluation of gravitational field is delicate because positive charges and mass are not independent in the field evaluation but interacting to each other as shown in Eqn. (3). In spite of this nonlinearity due to the source-fields interaction, the gravitational filed can be evaluated as long as the locality is assumed to be valid in physical field.

If the same amount of negative charges, instead of the positive charges, is put at the surface of radius $r$ in Fig. (3), the same effect should be expected for the gravitational field as before since the interaction strength of PVN charge to positive charge or negative charge is same in absolute value. Fig. (4) shows that gravitational field changes with the surface charges in Fig. (3). If there is no electric charge on the surface ($q_e$), gravitational field at a radial distance $d_r$ ($d_r \geq r$) depends only on the mass $M$, in which PVN charge ($-Q_\upsilon$) is given as $Q_\upsilon = |C_{pvn}| \cdot M$. If positive charges are increased – which can be induced naturally – or negative charges are enforced on the surface, gravitational field $g$ is reduced until the gravitational field effect by the mass ($g_m$) is disappeared.

In Fig. (4), $Q_\upsilon = 10^{-19} d_r^2 G^{-1} g_m$ ($M = d_r^2 G^{-1} g_m$); $q_e = d_r^2 k^{-1} E_\pm$; $q = 10^{-19} d_r^2 G^{-1} g$. If the $q_e$ is greater than $Q_\upsilon$ or less than $-Q_\upsilon$, gravitational field $g$ comes from the mass-charge interaction directly between the unit mass (kg) and the excess charges, $\pm(|q_e|-Q_\upsilon)$, respectively.

The dotted line in the Fig. (4) indicates electric field, $E = k(q_e - Q_v)/d_r^2$, by the PVN charge and physical charge $q_e$.

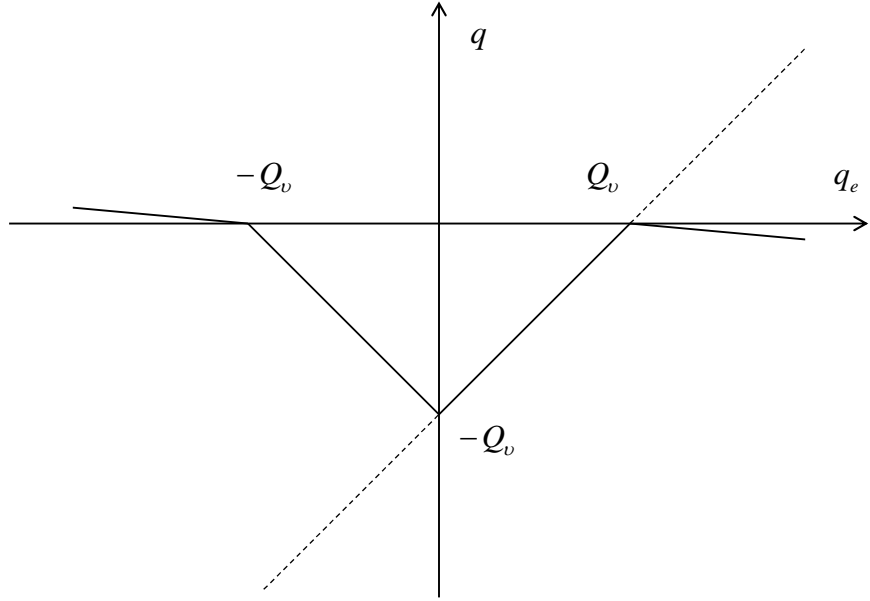

Figure 4. Gravitational field variation versus charges on the surface in Fig. (3)

Electrostatic field can be expressed by the source contributions as $\vec{E} = \vec{E}_+ + \vec{E}_- + \vec{E}_m$, in which $\vec{E}_+ = -\nabla\Phi_+$ from positive charges, $\vec{E}_- = -\nabla\Phi_-$ from negative charges, and $\vec{E}_m = -\nabla\Phi_m$ from the PVN charge for mass. Here, one critical assumption is imposed as $|\vec{E}_+|$ or $|\vec{E}_-|$ is not greater than $|\vec{E}_m|$ and $|\vec{E}_m| \neq 0$; otherwise, gravitational field intensity (N/kg) is order of $\sim 10^{-19} \times (|E_\pm| - |E_m|)$ in magnitude. It is for the mass-charge interaction when the electric charges are induced for the gravitational field by mass.

Because of the nonlinearity in mass-charge interaction, for instance, it can not be supposed that an arbitrary amount of positive charges and the same amount negative charges can be placed on a same position in the physical system, which can be assumed in classical electrostatics. Hence, let's sort out local electric field by its contribution for each component in a coordinate system as following:

If $(E_+ \cdot E_-) \geq 0$, then let's define $E_{g+} \equiv E_+$ and $E_{g-} \equiv E_-$; if $(E_+ \cdot E_-) < 0$ and $|E_+| \geq |E_-|$, then $E_{g+} \equiv E_+ + E_-$ and $E_{g-} \equiv 0$; if $(E_+ \cdot E_-) < 0$ and $|E_+| < |E_-|$, then $E_{g-} \equiv E_+ + E_-$ and $E_{g+} \equiv 0$. Now, if $(E_{g+} \cdot E_m) > 0$, $E_{g+} + E_m \Rightarrow -E_{g+} + E_m$ since positive charges are attractive to mass; otherwise, $E_{g+} + E_m \Rightarrow |E_{g+} + E_m| \cdot e_m$ due to the field interaction between $\vec{E}_{g+}$ and $\vec{E}_m$, but $|E_{g+} + E_m| \cdot e_m = E_{g+} + E_m$ since $|\vec{E}_+| \leq |\vec{E}_m|$ in which $e_m \equiv \frac{E_m}{|E_m|}$. However, $E_{g-} \Rightarrow -E_{g-}$ since mass is repulsive to negative charges. For each component the corrected field $E_c^i$ ($i = 1, 2, 3$) is expressed as

$$E_c^i \equiv \begin{cases} -E_{g+}^i + E_m^i - E_{g-}^i & : \text{ if } (E_{g+}^i \cdot E_m^i) \geq 0 \\ E_{g+}^i + E_m^i - E_{g-}^i & : \text{ otherwise} \end{cases} \quad (4)$$

$$= -(e_{g+}^i \cdot e_m^i) E_{g+}^i + E_m^i - E_{g-}^i \quad (5)$$

where $e_{sym}^i \equiv \frac{E_{sym}^i}{|E_{sym}^i|}$ (sym; g+, g−, m) and $e_m^i \neq 0$.

However, there is a correction factor ($f_{mc}$) from the relation between gravitational field and electrostatic field, which is came from the coupling strength ($C_{mc}$) of mass-charge interaction and given as

$$f_{mc} \sim \frac{k_c}{C_{mc}} \sim 0.7 \times 10^{-1} \quad (6)$$

With the correction factor each component of gravitational field $g^i$ in a physical space having mass and charges can be expressed as $g^i \sim f_{mc} E_c^i$. Therefore, the gravitational field $\vec{g}$ is expressed for each component as

$$g^i = f_{mc} \{-(e_{g+}^i \cdot e_m^i) E_{g+}^i + E_m^i - E_{g-}^i\} \quad (i = 1, 2, 3 \; ; e_m^i \neq 0) \quad (7)$$

or

$$\vec{g} = S(\vec{E}_{g+}, \vec{E}_m) - \vec{E}_{g-}$$

where $S(\vec{E}_{g+}, \vec{E}_m) = f_{mc}(\vec{E}_m \pm \vec{E}_{g+})$ with the sign from $\left( |\vec{E}_m + \vec{E}_{g+}|, |\vec{E}_m - \vec{E}_{g+}| \right)_{\text{smaller one}}$.

Gravitational field with mass-charge interaction is not linear due to the interaction of source fields. However, as long as locality, continuity, and smoothness of physical fields are assumed to be valid, gravitational field including mass-charge interaction can be evaluated as in Eqn. (7).

Inside conductor, for example, there is no electrostatic field for any outside stationary charge distribution, however; gravitational field from outside the conductor, is not changed inside the conductor as long as the conductor is isolated and electrically neutral because induced positive charges on the surface of conductor for the external PVN charge shield the gravitational field; however, the same amount of effect in the opposite direction is come from the induced negative charges on the other side.

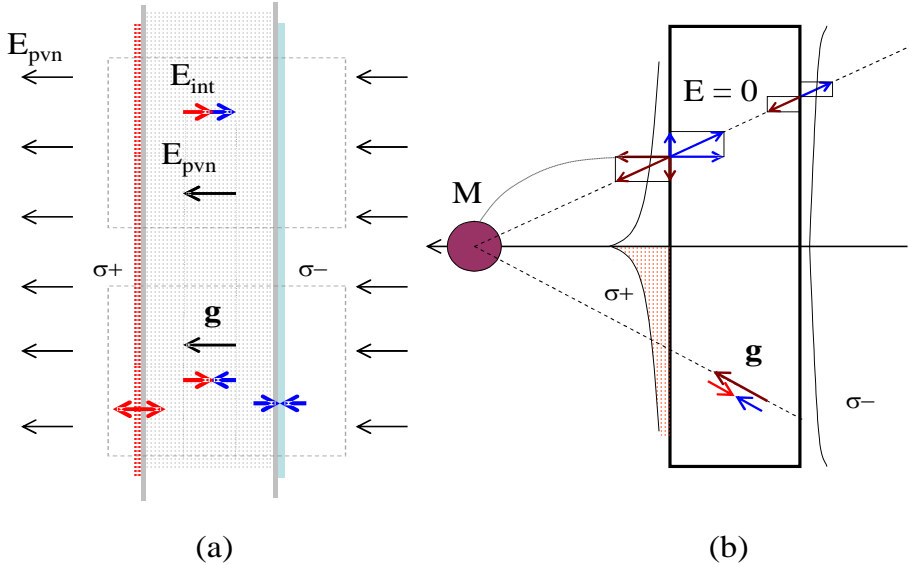

Figure 5. neutral conductors or conducting plates responding to external gravitational fields in cases : (a) source of the gravitational field is so far and (b) the source is near.

Fig. (5) shows (a) two parallel conducting plates that is infinitely extended and connected to each other and (b) conducting slab with nearby gravitational source $M$. If the total charge on the parallel plates in (a) is zero (neutral), for an external constant electric field ($\vec{E}_{pvn}$) the induced charge distribution on the plates is constant as $\sigma = |\sigma^+| = |\sigma^-|$, and the induced charge distribution produce the internal electric field ($\vec{E}_{int}$) to cancel out the external electric field. However, if the external electric field is from the PVN charge for mass object, gravitational field is not changed inside the plates; 50% of the gravitational field intensity is reduced by $\sigma^+$ distribution, but the same amount of effects is added in the opposite direction by $\sigma^-$ distribution. However, this reasoning as in case (a) can be generalized to case (b) because total charge in the conductor is zero.

As the positive charges in the conductor the negative charges also have been induced by the external field. This means that the internal electric field $E_{int}$ is produced by the same amount of contribution from the positive charges and the negative charges. Hence, gravitational field inside the conductor in case (b) is not changed as in the case (a). If the negative charge distribution is not changed by the external field in a spherical symmetric geometry as in Fig. (3), in which electric field inside the conductor is zero, the internal electric field $E_{int}$ is produced by the positive charges only.

If the total charge in a conductor is not zero, gravitational field inside and outside the conductor can be changed. The change inside the conductor is the result of interactions with the charges in the conductor; hence, the effect of change can be appeared as an internal force. However, the change of gravitational field outside the conductor should be appeared as an inertial mass change of the conductor for external field. Although it has been still augmentable and needs more detail investigation, the interaction of gravitational field with electrostatic charges has been studied in phenomenological point of view [8].

### 3.2 Maxwell's equations with mass

In classical electromagnetism, Maxwell's equations describe the time dependent relation between electric field $\vec{E}$ (electric displacement $\vec{D} = \varepsilon \vec{E}$) and magnetic induction $\vec{B}$ (magnetic field $\vec{H} = \mu^{-1}\vec{B}$). If the PVN charge for mass is included as a source of the electric field $\vec{E}$, magnetic induction $\vec{B}$ also can be generated with time dependent $\vec{E}$. As a simple case, if mass is only source of the electric field in free space,

$$\nabla \cdot \vec{E} = \frac{\rho_{pvn}}{\varepsilon_o}$$

$$\nabla \cdot \vec{B} = 0$$

$$\nabla \times \vec{E} + \frac{\partial \vec{B}}{\partial t} = 0 \quad (8)$$

$$\nabla \times \vec{B} - \frac{1}{c^2}\frac{\partial \vec{E}}{\partial t} = \mu_o \vec{J}_{pvn} \qquad \left(c = \frac{1}{\sqrt{\varepsilon_0 \mu_0}}\right)$$

where c is light velocity in free space, and $\rho_{pvn}[C/m^3] \sim C_{pvn} \cdot \rho_m \ [kg/m^3]$. To satisfy the continuity equation as $\nabla \cdot \vec{J}_{pvn} = -\frac{\partial \rho_{pvn}}{\partial t}$ the current of PVN charge or simply mass current $\vec{J}_{pvn}$ should be included, which means that magnetic induction $\vec{B}$ can be generated by mass flow. However, it should be distinguished from the frame-dragging effects in general relativity [9].

## 4. Antigravity

In spite of the fact that antigravity phenomena have been known [10], it has not been accepted in physics because any proper explanation couldn't be found with contemporary theories in physics. To get a quantitative explanation about the antigravity phenomena the mechanical properties and thermal properties of the confined vacuum particles in the Imaginary Space should be known. However, a qualitative explanation is possible. Considering that gravitational field and electromagnetic field are originated with unique mechanism in the 4-D complex space, two distinctive fields in phenomenology can interact to each other.

### 4.1 Parallel Plate Capacitor

On a physical object -- mass or charge -- introduced in the physical space, the reaction of the confined vacuum particles in the Imaginary Space is to get new equilibrium in the space. In other words, vacuum particles try to move from high density region to low density region for net-mass density and net-charge density.

As a simple and ideal case, let us assume that electric field does not exist outside capacitor but exist inside uniformly. Thus, the outside is almost free space except gravitational fields. If the elastic property of vacuum particles is corresponded to bundles of strings – on each of which vacuum particles are connected, the number density of vacuum particles inside of capacitor is smaller than the number density outside because the difference of number density is appeared as the electric force between the capacitor plates. The string tension inside is getting higher with increasing electric potential energy. The electric potential energy density in the capacitor, $u_e$ is related as

$$\frac{F_e}{S} = u_e = \frac{1}{2}\sigma E \Leftrightarrow \frac{1}{n} \tag{8}$$

Here, $F_e/S$ is electric force per unit area, $\sigma$ is surface charge density, and $n$ is volume number density of vacuum particle. Therefore, the corresponding distribution of vacuum particles inside the capacitor is similar to the distribution made by a negative massive object.

Let us suppose that the number density of vacuum particles is reduced along the direction of electric field ($\vec{E}$). Then, let us think one string of vacuum particles, which is connected between the two parallel plates of capacitor. With increasing $\vec{E}$, the tension of the string is also increased, and the number density of vacuum particles in the string is reduced. If the volume number density of vacuum particles is reduced with a ratio $x$ as $n_x = \frac{n_g}{x}$, the corresponding tension is increased as $T_x = xT_g$. Here, $T_g$ is the string tension and $n_g$ is volume number density under gravity.

The difference of tension between outside and inside capacitor result in the electric force between the capacitor plates. Thus, $u_e = n_s(x-1)T_g$ with Eqn. (8). Here, $n_s$ is string number density on the surface that is parallel to the capacitor plates.

$$n_x = n_g \left( \frac{1}{1 + \frac{u_e}{n_s T_g}} \right) \tag{9}$$

However, $T_g = kd_g$ ($k$ : spring constant, $d_g$ : separation distance of vacuum particles under gravity).

If $d_0$ is the separation distance of vacuum particles and $T_0$ is the string tension in free space, $T_g = \left( \frac{d_g}{d_0^2} \right) \cdot m_e c^2$ since $k = \frac{m_e c^2}{d_0^2}$ [1]. If the number density difference is defined as $\Delta n \equiv n_g - n_x$,

$$\Delta n = n_g \left( \frac{\frac{d_0^2}{n_s d_g} \left( \frac{u_e}{m_e c^2} \right)}{1 + \frac{d_0^2}{n_s d_g} \left( \frac{u_e}{m_e c^2} \right)} \right)$$

$$\approx n_g \cdot \frac{d_0^2}{n_s d_g} \frac{u_e}{m_e c^2} . \tag{10}$$

For the order of estimation, if a simple cubic structure is used for the distribution of vacuum particles; $n_0 \sim \frac{1}{d_0^3}$, $n_g \sim \frac{1}{d_0^2 d_g}$, and $n_s \sim \frac{1}{d_0^2}$. Hence,

$$\Delta n \approx \frac{u_e}{m_e c^2} \left( \frac{d_0^2}{d_g^2} \right) . \tag{11}$$

However, it is estimated as $n_g \approx n_0$ and $d_g \approx d_0$. The reason is as following. Since gravitational field is compared with a PVN charge $Q_v$, let us think a spherical conducting shell with radius $r$, and charge $Q_v$ is uniformly distributed on surface of the shell. Inside the conducting shell, $n_0$ is corresponded; outside, $n_g$ is corresponded. The pressure on the conducting shell is come from the difference of tension in vacuum particle strings inside and outside of the shell as

$$p = \frac{Q_v^2}{32\pi^2 \varepsilon_0 r^4} \tag{12}$$

$$= n_s \Delta T$$

$$= n_s k(d_0 - d_g). \tag{13}$$

Here, $k$ is spring constant; $d_0$ is separation distance of vacuum particles in free space; and, $d_g$ is separation distance under gravity at the radius $r$. From Eqn.(12) and Eqn.(13),

$$\frac{Q_v^2}{32\pi^2 \varepsilon_0 r^4} = \frac{m_e c^2}{d_0^3}\left(1 - \frac{d_g}{d_0}\right) \tag{14}$$

since $k = \frac{m_e c^2}{d_0^2}$.

If the parallel plate capacitor is on the earth, $Q_v \approx -6 \times 10^5 \text{C}$ with Eqn. (1) and the radius of the earth, $r \approx 6 \times 10^6 \text{m}$. From Eqn. (14),

$$\frac{d_g}{d_0} \approx 1 - 10^{12} \cdot d_0^3,$$

where, $d_0 \sim 10^{-23} \text{m}$ [11]. Thus, $d_g \sim d_0$, and $n_0 \sim n_g$.

Therefore, Eqn.(11) can be expressed as

$$\Delta n \approx \frac{u_e}{m_e c^2}. \tag{15}$$

Some vacuum particles are pushed out of the capacitor as a reaction to the electric potential energy supplied. Since each vacuum particle in free space is confined with $\varepsilon_v \approx -m_e c^2$, the number of expelled vacuum particles is directly proportional to the electric potential energy.

Inside the capacitor is in new equilibrium with external electric fields, but it is not in equilibrium with gravitational fields. Since the vacuum particle number density has been reduced inside of the capacitor, the capacitor is corresponded to (or simply contains) less negative mass in physical vacuum, which is similar to the case of negative mass object. Therefore, the capacitor has antigravity effects. It is not easy to assign a PVN mass for this case, but a buoyant force can be estimated as

$$\Delta M \approx \frac{u_e}{c^2} V \tag{16}$$

Here, $\Delta M$ is total mass difference in physical vacuum, which is compared with outside of the capacitor with same volume of space ($V$). On the surface of the earth gravitational acceleration $g$ is almost constant as $g = 9.8 \text{ (m/s}^2\text{)}$. Then, the buoyant force is

$$F_b \approx \frac{U_e}{c^2} g. \qquad (17)$$

That is

$$F_b[N] \approx 10^{-16} U_e[J]. \qquad (18)$$

Indeed, there is the antigravity effect, but it is so small to be measured in the case of parallel plate capacitor.

## 5. **Electric and Magnetic Fields of the Earth**

### 5.1 Natural electric field of the earth

It has been known that the earth's natural electric field exists, and the direction of the field is pointing to the center of the earth. The intensity of the electric field varies with atmospheric conditions and also depends on places and the time of day for the measurements, but the maximum order of magnitude is 2 as $E \approx 100 \sim 200$ (N/C) [12].

If the primitive-virtual negative (PVN) charge is estimated for the earth, the PVN charge $Q_\nu$ is $\sim -6 \times 10^5$ C with the PVN constant ($C_{pvn} \equiv -10^{-19}$ [C/kg]) in Eqn. (1), and the earth's mass ($\sim 6 \times 10^{24}$ kg). Since the radius of the earth is about 6378 km, the electric field on the earth's surface is estimated as $E \sim 130$ (N/C) with pointing downward. It is consistent with the measurements in the order of magnitude. In fact, the electric field on the earth is determined not only by the PVN charge of the earth but also by the charge distributions inside the earth and in atmosphere.

In the view of electrical property the earth's atmosphere can be divided by planetary boundary layer (PBL) up to a few km above the ground, low atmosphere, ionosphere (70-1500 km), and magnetosphere. The ionosphere is divided by D, E, F1, and F2 layer, and the main ionization source is the solar UV radiation and cosmic rays. The diurnal variations in total electron number density (TEC) and electrical conductivity in ionosphere are affected mainly by the solar radiation.

The average life time for the ionized particles depends on the recombination process and chemical composition in each layer; hence, the average life time in E-layer and F-layer, for example, is ~30 sec and a few hours, respectively. However, electrical relaxation time ($\tau$), which is directly proportional to the electrical conductivity, is $\sim 10^{-4}$ sec at 70 km in ionosphere and increasing with height. In low atmosphere, $\tau$ is 4 sec at 18 km and 5-40 min at 10 m above the ground. On the other hand, the relaxation time on the earth's surface is $\sim 10^{-5}$ sec [13].

Owing to the PVN charges of the sun, the earth, and other planets in solar system, the electric equilibrium state need to be considered prior to the dynamic state and the magnetic effect in MHD approximation. It can be supposed that the earth is immersed in a very thin plasma sea under the interplanetary magnetic field (IMF) in space. Once the sun's PVN effect is considered, some positive charges from the outer space should be induced on the sunny side of the earth and negative charges are pushed back further alongside the gravitational neutral surface between the sun and the earth. However, it is desirable to consider the charge distribution inside the earth's system because the conductivity in the interplanetary space is much lower than the earth's ionosphere.

If there is any excess charge distribution in the atmosphere, the excess charge distribution should move to the lowest boundary of the atmosphere or to outer space. If the conductivity in the interplanetary space is ignored, the outward excess charge distribution should be at the top of the ionosphere. These excess charge distributions at the lowest boundary of the atmosphere and the top of the ionosphere should be distorted due to the solar PVN effect.

To make a simple approach, let's suppose that mainly the ionosphere of the earth is affected by the PVN charges from outer space and the earth itself. In the view of the electrostatic equilibrium, the lowest boundary of the atmosphere can be set as the low boundary of ionosphere in that the conductivity of PBL is so low and close to an insulator. However, the electric field in low atmosphere is dependent to PBL charge distribution – electrode effect -- and the telluric charge distribution, in which the telluric charge distribution is affected by the ionospheric charge distribution to minimize the external PVN effects. Here, the telluric charges are the source of telluric current [14] and assumed as mostly electrons.

Although the effect of the solar PVN on the earth is small compared with the effect of earth's PVN, the amount of induced charges on the top of the ionosphere -- boundary between the earth and the outer space -- is not small but even bigger than the earth's PVN.

Using the known data (solar mass : $\sim 2.0 \times 10^{30}$ kg, the distance from the earth : $\sim 1.5 \times 10^{11}$ m, and the earth's radius : $\sim 6.4 \times 10^{6}$ m) if the electric shielding effect by the interstellar space plasma is ignored, the total amount of induced charges on the top of the ionosphere can be reached up to $8.5 \times 10^{6}$ C on the top of the ionosphere. It can be supposed that the amount of charges induced by the solar PVN on the earth is big enough to distort the spherical symmetric charge distribution by the earth's PVN. Here, the effect of the moon on the earth is ignorable since the gravitational field intensity by the moon is about 2 orders smaller than the sun.

If two sides of the earth – dayside and night side – are compared, at the top of the ionosphere there should be a net excess positive charge distribution on dayside and a net excess negative charge distribution on night side. However, the earth's ionosphere is far from a perfect conductor but can be assumed as an ohmic conductor. At the low boundary of the ionosphere, on the other hand, there should be a net excess positive charge distribution on night side.

On dayside in the low boundary of ionosphere; however, there should be a net excess negative charge distribution that is corresponding to the positive charge distributions in PBL (including telluric charge distribution) and at the top of the ionosphere. If the electric field intensity at the base of thermosphere is $E \sim 10^{-6}$ V/m [15] with downward direction, a net excess positive charge distribution can be created at the bottom of the low boundary of ionosphere with the net excess negative charge distribution at the top. Furthermore, the atmospheric height of the low boundary of ionosphere can be variable according to the net excess charge distributions on dayside ionosphere.

It can be said, at least, that TEC at night side in the low boundary of ionosphere is lower than the TEC at dayside.

In the view from the outer space (or in heliocentric coordinate system), the earth's atmosphere is rotating with the earth and, thus, the electric charge distribution in the atmosphere keeps being changed during the day and night due to the solar PVN.

In the low atmosphere including PBL, the variation of charge density is small during the day due to the low conductivity. However, telluric charges respond for the ionospheric variation instead of the low atmosphere. This means that the solar PVN affect not only on the charge distribution in the low boundary and the upper boundary of ionosphere but also on the telluric charge distribution of the earth.

To minimize the solar PVN effect in the low atmosphere the telluric charges are attracted toward the sunny side. Even though the electrical conductivity is so low in the low atmosphere, electrical conduction process is taken slowly from the low boundary of ionosphere, PBL, and to the surface of the earth, in which the conduction current is $\sim 10^{-12} \text{ A}/\text{m}^2$.

Hence, there should be a net excess positive charge distribution on the night side at the low boundary of ionosphere of the earth, and there should be relatively more telluric charges on dayside than night side. From the facts that D and E layers are disappeared and that F1 and F2 layers are combined at night, the low boundary of ionosphere can be thought as the D-layer and some part of E-layer.

On the dayside of the earth, electrical conductivity is increased in upper atmosphere due to the solar radiation (UV) and in the low atmosphere mainly by the temperature increments in the air and of the earth's surface. For instant, the conductivity at 300 km above ground starts increasing about one hour before the sunrise on the ground, and positive charges keep being induced to the top of ionosphere.

Nearby the zero degree of solar zenith angle; the electrical conductivity should be maximized. The positive charge density has been increased at the top of ionosphere; then, the negative charge density at the low boundary of ionosphere and the telluric charge density become to lower than in the morning. Furthermore, these charge distributions are moving from East to West like steady currents in an earth-fixed coordinate system. It results in a negative charge accumulation before the solar noon at the low boundary of ionosphere due to the electrical conductivity that is getting smaller from East to West. The telluric charges respond for the variation of charge density in the low boundary of ionosphere, and build up right below the charge accumulation in the low boundary of ionosphere. Hence, electric field on the surface of the earth has a peak before the solar noon although it depends on latitudes and seasons of the earth.

On night side of the earth; on the other hand, there should be the net excess positive charge distribution in the low boundary of ionosphere and relatively less telluric charge density compared to dayside. However, the charge density at the PBL and the electrical conductivity in the atmosphere keep being reduced during night due to the temperature drop and no direct solar radiation; thus, the telluric charge density is also reduced, and the electric field on the surface of the earth is getting reduced at night.

The charge distribution at the low boundary of ionosphere is shifted to the East due to the earth's rotation and the reduced conductivity at night. This means that there should be relatively more negative charges on West side and more telluric charges induced to the West. It can make another, but small, peak of the electric field on the earth's surface.

In the view from an earth-fixed coordinate system these charge distributions keep moving from East to West. With the Coriolis effect in the rotating system, diurnal variations of ionosphere, earth's magnetic field, and telluric currents also should be correlated to the diurnal variation of the earth's natural electric field.

Although above qualitative reasoning cannot be enough to explain for the diurnal and seasonal variations of the earth's natural electric field because many physical effects should be considered together even in a fair weather and solar quite (Sq) condition, the general pattern of diurnal variation can be explained simply by considering of the electric equilibrium with PVN charges of the sun and the earth.

## 5.2 Geomagnetic field

To explain the origin of the earth's main magnetic field that is approximately a magnetic dipole, there have been many theories suggested. Especially, dynamo theory [16] seems to have been concerned much; however, it is very difficult to understand the theory itself, of course, in person. The theory needs a seed magnetic field to initiate whole mechanism that makes electric currents flowing around the core of the earth.

Even though the seed magnetic field in the dynamo theory is possible without an external source owing to the earth's rotation with introducing the PVN charges of the earth, in which the density of charge should be proportional to the density of mass, another alternative mechanism can be speculated as following; however, a detail investigation is still remained open.

At the beginning, a large swirling nebula had its own PVN charge $Q_\upsilon$ and angular momentum energy; hence, there was a primordial magnetic field. With the magnetic field the swirling charged particles in plasma state were getting separated -- dynamo process, in which positive ions were attracted to the center of the magnetic field and negative charges – mostly electrons – were pushed outward. The mass density at the center of the magnetic field was getting increased, and the process of charge separation was slow but continued.

The density of negative charge distribution at outmost layer of the primordial earth should be proportional to $\sigma_{\mathrm{pvn}} \equiv Q_\upsilon / 4\pi\, r_\oplus^2$, which is the source of seed magnetic field. If an increment factor is given as $f(M_\oplus, r_\oplus, \omega, \varepsilon_\mathrm{r}, \mu_\mathrm{r},,,)$ from the relation of force equilibrium between Lorentz force and electrostatic force, the induced charge density $\sigma_-^{ind} \sim \sigma_{pnv} f$. The induced charge distribution should be an oblate shape as shown in Fig. (6) due to the direction of Lorentz force. Hence, it is supposed that $\sigma_-^{ind}(\theta) \sim \sigma_{pnv} f(M_\oplus, r_\oplus, \omega, \varepsilon_r, \mu_r,,,)\sin^2(\theta)$. Here, $M_\oplus$ is the earth's mass; $r_\oplus$, the radius of primordial earth; $\omega$, angular speed of the earth's rotation; $\varepsilon_r$ and $\mu_r$ are relative permittivity and relative permeability inside the earth, respectively. Also, many other variables in magnetohydrodynamic (MHD) should be included in the function of increment factor.

Again, these additional induced charges produced additional magnetic field; in turn, the additional magnetic field generated another induced charges. As long as Ohmic loss was overcome and electric conductivity $\sigma$ was not zero, the process of charge separation was continued until the earth had been cooled down. The variables, such as $\sigma_{pvn,} r_\oplus, \omega, \varepsilon_r, \mu_r$, etc., are time-dependent in that the process of geomagnetic field creation should had taken a long time. If the increment factor is small enough to make an approximation, total induced charges can be expressed as $\sigma_-(\theta) \sim \sigma_o \sin^2(\theta)$, in which

$$\sigma_o \sim \sigma_{pnv} \exp\left(\sum_i^N f_i\right). \tag{19}$$

$N$ is the number of iteration process, and $\sigma_{pvn} \sim Q_\upsilon / 4\pi R_\oplus^2$ at present earth. Even though a detail investigation using magnetohydrodynamics (MHD) is still remained open, it shows how the charge separation is possible during the period of planet formation in Eqn. (19). Once geomagnetic field was created, it should have been maintained with environmental situation.

It has been known that the earth has layers, such as inner core, outer core, mantle, and crust. Let us suppose that electrical ionization is possible in liquid state regions, outer core and some region in lower mantle. Then, positive ions should have a distribution from outer core to lower mantle since the positive ions are attractive to the center of the earth. However, negative ions (mainly electrons) moved outward; became to form a narrow distribution, for example, in Moho (Mohorovicic Discontinuity) layer between upper mantle and the crust even though how the electrons could move through the solidified and nonconductive materials might be answered in the history of the earth.

Like branches of a tree, many conductive passages can be formed from the negative charge distribution to near the earth's surface and also to the core region. Hence, this induced negative charge distribution can be interpreted as one of the sources for telluric currents.

Due to the earth's rotation and the distribution of PVN charge of the earth, magnetic dipole moment exist as $\mathsf{m}_{pvn} \sim Q_\upsilon R_\oplus^2 / T \sim 10^{14} \,(\mathrm{A} \cdot \mathrm{m}^2)$ with pointing geographic south pole, in which $Q_\upsilon \sim \mathrm{C}_{pvn} M_\oplus$, $R_\oplus$ is the radius of earth at present, and T is the period of rotation. As long as the earth's rotational speed is not changed, the magnetic field from the dipole moment can be considered as like a constant external magnetic field.

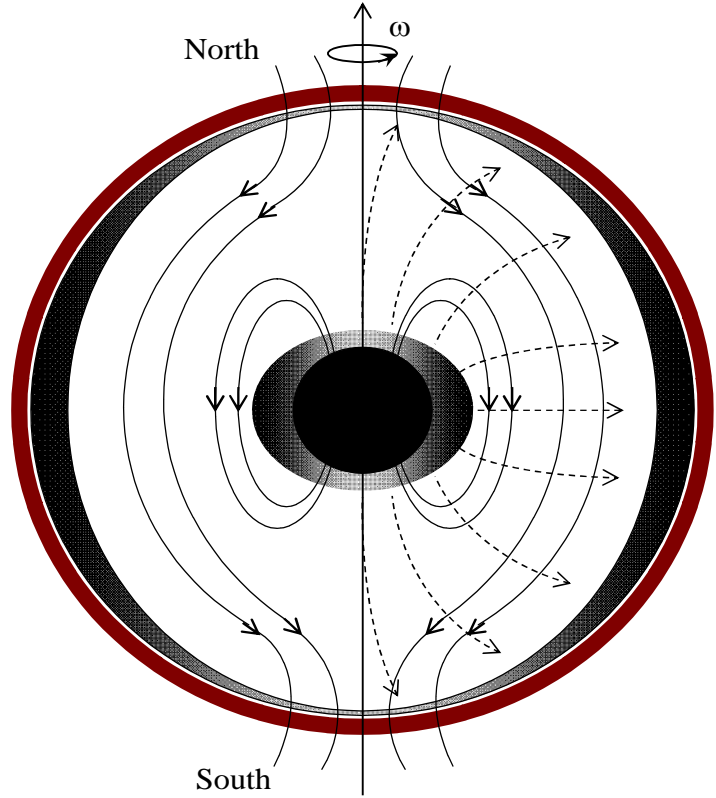

Figure (6) Magnetic and electric fields inside the earth

From the primordial dynamo process, if the earth has induced charges $+Q_\oplus$ above inner core and $-Q_\oplus$ below the crust, these charge distribution should have been maintained with geomagnetic field of the earth. As shown in Fig. (6), the induced charge distributions are not spherical but rather be oblate to make the balance with Lorentz force. Two electric quadrupole moments become to exist in the positive charge distribution at around the outer core (red) and negative charge distribution below the crust (blue) as shown in Fig. (6) that might be the earth's geomagnetic fields after the primordial process.

The dotted lines indicate electric fields only by the induced charge distributions. If the magnitudes of two quadrupole moments are same except the signs, electric fields in the atmosphere of the earth (outside) is ascribed to only the PVN charge.

In the view from outside the earth (inertial frame), rotating two quadrupole moments generate two magnetic dipole moments; however, the angular momentum energy of the primordial earth should had been reduced due to a backward torque, which is the reaction against the charge separation and the magnetic field generation inside and outside the earth. If the angular momentum energy of the earth was not big enough to overcome the backward torque, the rotation of the earth should have been disappeared due to the energy dissipation through ohmic loss and thermal energy generated in the medium inside the earth when the magnetic field was generated.

Once the magnetic fields of the earth has been stabilized and time-independent in an inertial frame (non-rotating frame); then, $\nabla \times \vec{E} = 0$. Since the magnetic field can be assumed that it has axial symmetry that means $\nabla_\varphi \vec{B}(r,\theta,\varphi) = 0$, the same amount of magnetic field strength should be measured on the earth's surface (rotating frame).

The gravitational field inside the earth – from the outer core to under the crust -- is also affected by the induced charges since the amount of induced charges are much bigger than the PVN charge $Q_\upsilon$. As shown in Fig. (4), the gravitational field should be proportional to the electric filed intensity as $\vec{g} \sim C_{pvn} \times \vec{E}$, which means that gravitational force inside the earth is practically ignorable. Instead, thermal and magnetic pressures should be the dominant factors to maintain the equilibrium state inside the earth.

Since two magnetic field sources exist inside the earth with different directions, in which one is pointing geographical north and the other is pointing geographical south at the center of the earth as shown in Fig. (6), the magnetic pressure pushes up the earth's crust and pushes down the core region. Hence, the rotation of the core region and its translational oscillations are getting more flexible in that the core region become isolated from surrounding medium due to the magnetic pressure. If there is an unbalanced force inside the earth, the rotational axis of the core region can be tilted to get a more stable state as of today. In that case, the rotation of the earth gets faster owing to the angular momentum conservation; then, it makes more magnetic pressure on the core region and leads to another metastable state.

As shown in Fig. (6), there are two magnetic dipole moments, one of which is from positive charge distribution in outer core (inner part); the other, from negative charge distribution below the earth crust (outer part). If the charge distribution in the outer part is given as $\sigma_-(\theta) \sim \sigma_o \sin^2(\theta)$, the magnetic dipole moment $m_{outer} \sim \frac{1}{2} \omega Q_\oplus R_\oplus^2$, in which $Q_\oplus = \frac{8\pi}{3} R_\oplus^2 \sigma_o$.

Once geomagnetic field was created and stabilized, the magnetic field and the earth's rotational period can be assumed as constants; then, the induced charge $Q_\oplus$ should be proportional to $Q_\upsilon$ and $R_\oplus$ because $Q_\upsilon$ is the initiating factor generating the primordial magnetic field and $R_\oplus$ is related to the velocity term in the expression of Lorentz force. Therefore, we can set as $Q_\oplus = (R_\oplus/R_{ref}) Q_\upsilon$ with a constant $R_{ref}$. Here, $R_{ref}$ can be defined as a limit on which the effect of Lorentz force is ignorable and only electrostatic force is effective. Hence, $m_{outer} \sim \frac{1}{2} R_{ref}^{-1} \omega Q_\upsilon R_\oplus^3$ with pointing south. On the other hand, $m_{inner} \sim -(R_{core}/R_\oplus)^3 m_{outer}$ with pointing north, which means 3 orders smaller than $m_{outer}$. Using the PVN constant in Eqn. (1), the magnetic dipole moment of the earth is estimated as

$$m_\oplus \sim \left(\pi \times 10^{-19}/R_{ref}\right)\frac{M_\oplus R_\oplus^3}{T_\oplus} \qquad (20)$$

with pointing south. The proportional constant in Eqn. (20) has the dimension of ampere *per* unit mechanical momentum. It shows that the magnetic dipole moment is proportional to mass, volume, and angular velocity of the earth.

The constant parameter $R_{ref}$ in Eqn.(20) should be related with thermodynamic parameters inside the earth; thus, it can be different for each planet. Nevertheless, let's suppose $R_{ref}$ to be a unit length as a matter of convenience.

With known data about the earth, such as $R_\oplus \approx 6.38 \times 10^6$ m; inner core radius [17]; $T = 86400$ sec; $M_\oplus \approx 5.97 \times 10^{24}$ kg, the earth's magnetic dipole moment $m_\oplus \sim 5.6 \times 10^{21}\,(A \cdot m^2)$. It is about 1 order smaller than the data [18] reported as $m_\oplus \sim 8 \times 10^{22}\,(A \cdot m^2)$.

In Fig. (7), magnetic dipole moments for planets in solar system are compared with data [18]. The primordial magnetic dipole moments (PMD in the figure) has a similar pattern to the data (NASA) with ratios from $10^7$ to $10^8$. The ratio of data (NASA) to the estimated magnetic dipole moments (EMD) is shown in (RATIO) in the figure.

Since the data (NASA) should be based on magnetic field measurements around planets, the effect of magnetic susceptibility for ferromagnetic materials -- for example, iron–rich materials inside the planet -- should be considered as an enhancement factor of the EMD. If the temperature inside the planet is above Curie temperature, $T_c$, for instance, $T_c = 1043$ K ( 770 $^\circ$C ) for iron, it is not possible to expect the enhancement factor. As a reference, the Moho temperature of the earth has been estimated above or close to the Curie temperature. Hence, some iron-rich materials only in the crust of the earth contribute to the enhancement factor. The estimation for the planet Mercury needs 2 order of the enhancement factor. Since the iron core of mercury contains $50\% - 75\%$ of the planet mass and it is even molten [19], more iron–rich materials in the crust of Mercury than of the earth can be expected and/or simply the temperature inside the mercury can be lower than the earth.

In contrast, the ratio (RATIO) is less than one for Jovian planets, such as Jupiter, Saturn, Uranus, and Neptune in the Fig. (7) . It implies that the interior structure and composition of Jovian planets (or gas giants) should be different from terrestrial planets (like Earth). First, Jovian planet has not a solid surface but liquid or gas with decreasing density, and it has been known that the planet's interior consists of H, He, and H compounds. The magnetic field direction of Jovian planets is aligned in the reverse direction of the Earth's.

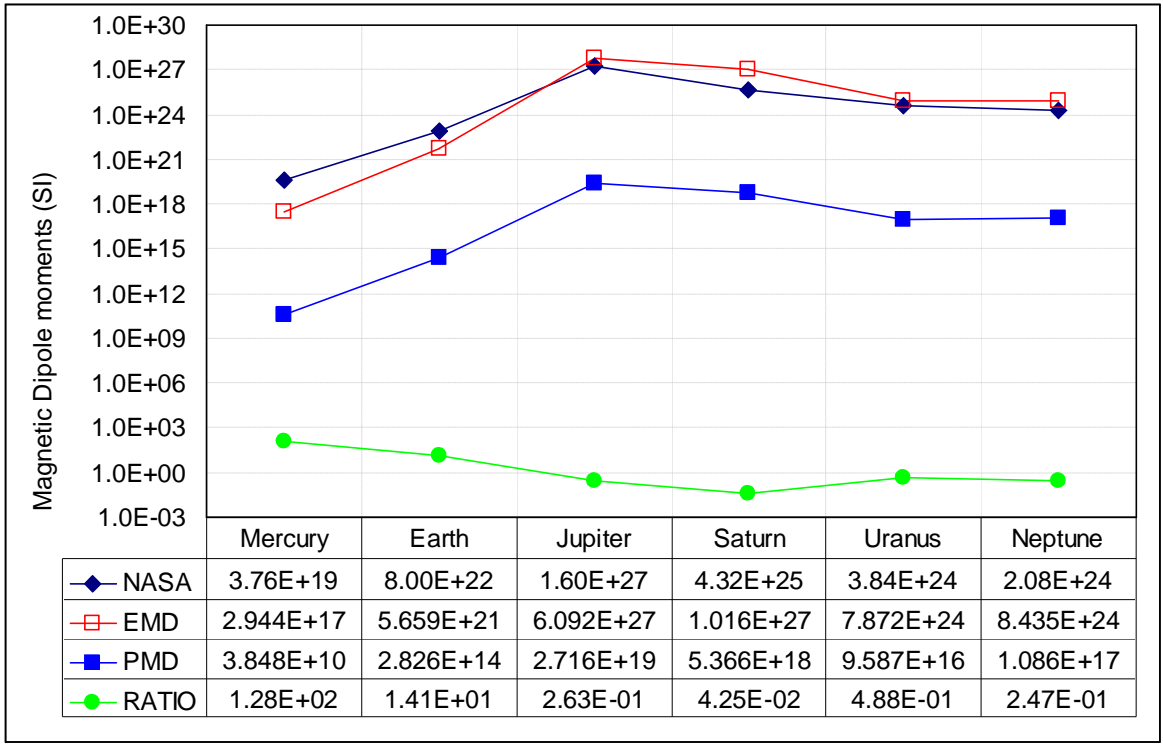

Figure (7) Magnetic dipole moments of planets

It has been suggested in theories and experiments [20] as following:
There should be a liquid metal hydrogen layer, which is an exotic state of hydrogen under extremely high pressure inside Jovian planets. The liquid metal hydrogen layer is extended from core (if any) to liquid molecular hydrogen layer with a transition zone (if possible). The liquid molecular hydrogen layer is considered as an insulator, but liquid metal hydrogen layer is supposed to have superconductivity.

If the interior composition of the Jovian planets is made of the liquid metal hydrogen layer and the liquid molecular hydrogen layer, the charge separation is initiated by the primordial magnetic field. However, the separated or detached electrons from the liquid metal hydrogen layer have no contribution to the magnetic field increment because the resistance of superconductor is zero; hence, the electrons cannot follow the planet rotation. Only positive charges in the liquid metal hydrogen layer contribute to the increment of planet's magnetic field. Hence, the ratio of NASA to EMD for Jovian planets means that the liquid metal hydrogen layer should be extended up to $0.64\,R_J$ in Jupiter and $0.35\,R_S$ in Saturn, for instance.

## 6. Earth's Gravity and Electric Equilibrium

In the view of 4-D complex space, gravitational field and electric field are not immutable by each other but can be changed through the interaction with each other. In other words, a spatial and time variation of gravitational field is related with the corresponding spatial and time variation of electric field; the variation of electric field is also related to the variation of magnetic field in the space. Therefore, gravitational field, electric field, and magnetic field in physical space, they are correlated to one another.

The earth is in electrical equilibrium state between inside the earth and outer space – mainly the sun – through the atmosphere. The equilibrium is supposed to be maintained slowly through the atmosphere and the earth's surface or abruptly, sometimes, through sudden activities such as earthquakes, volcano eruptions, severe thunder storms, etc. It has been known that natural phenomena in atmosphere and inside the earth have relations with electric field variations and/or magnetic field variations [21] on the earth, and the Sun's activities [22].

### 6.1 Gravitational anomalies

Local gravitational field on the earth is affected by the external gravity sources, such as the sun, moon, etc. As like the interior charges of a conductor, the ionospheric and telluric charges respond for the external gravitational field. However, the ionosphere of the earth is not a perfect conductor; although it is a minor fluctuation, local gravitational field on the earth is also affected by transient effects of telluric (from Moho layer and up to the surface), planetary boundary layer (PBL), and ionospheric charge distributions. One example is shown in the measurement of gravitational constant G [23], in which the constant G cannot be measured with high accuracy. In other words, the constant G cannot be determined more than 3 digits after decimal in the measurement.

Furthermore, if an abruptly enforced local disturbance, for example, is occurred in atmosphere, the response is different in each layer of ionosphere and from telluric charges due to the different electrical conductivities [13]. These different responses can result to a temporal gravitational anomaly in local area on the earth.

Even though the magnitude of gravity variations during solar eclipses is so small ($\sim 10^{-8}$ m/s$^2$), the variation of gravity is caused by the solar eclipse and it has not been explained clearly last half century [24]. During the solar eclipse, not only a sudden decrease of the strength of vertical gravity (10-12 µgal [25], 5-7 µgal [26], 1µgal $= 10^{-8}$ m/s$^2$) is observed before the first contact and after the fourth contact, but also the tilt of the apparent vertical direction is observed [27].

Since the moon blocks the solar radiation, it have been reported that total electron content (TEC) in ionosphere is reduced as much as 20%-50% [28]. As a correlation to the changes in ionosphere, the variations of atmospheric electromagnetic field also have been reported [29]. It has been known that the electric conductivity nearby the earth's surface is increased on the eclipse region, while electric potential gradient is reduced.

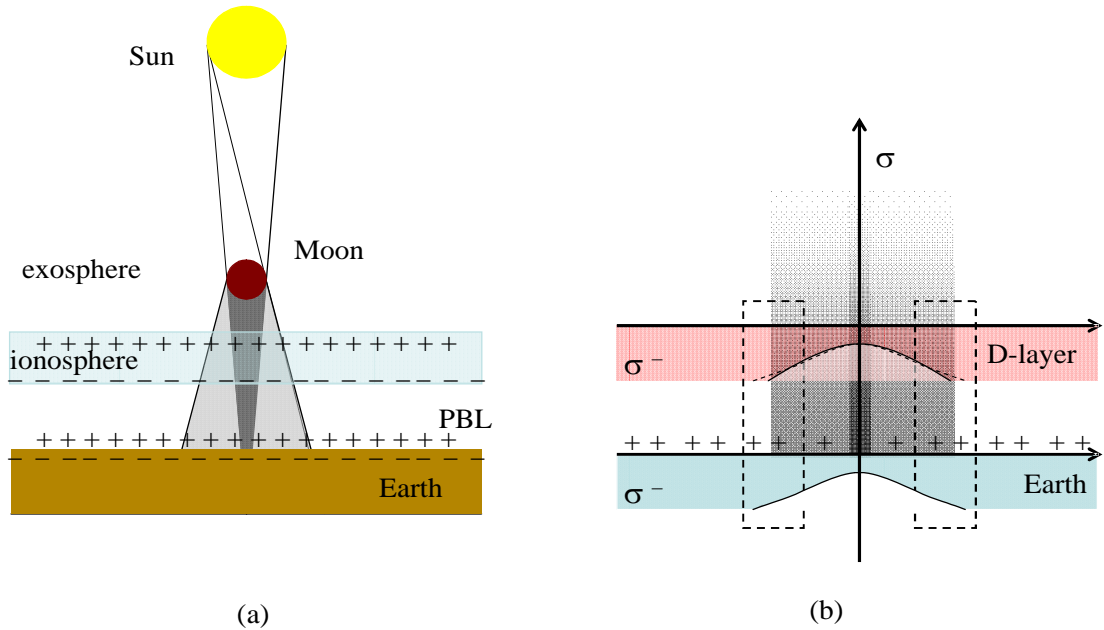

Figure (8) : (a) induced charges in ionosphere, PBL, and underground for external gravitations; and (b) the schematic drawing of the charge distributions during the solar eclipse.

Although it is still based on a theoretical speculation, the gravitational anomaly during the solar eclipse can be explained qualitatively as following: Fig. (8-a) is for the schematic drawing for the solar eclipse with assuming that the ionosphere and the telluric charge distribution can be approximated using conducting parallel plates. It shows net excess positive charge distribution in the upper boundary of ionosphere and net excess negative charge distribution in the low boundary. There are positive charge distribution in PBL and the telluric charge distribution underground.

The amount of induced positive charges by the lunar PVN is $\sim 10^2$ C to the lunar side in the upper boundary of ionosphere, in which the amount of induced charges by the lunar PVN is about 4 orders smaller than the solar PVN. Hence, the gravitational anomaly should be caused not by the lunar PVN effect but by the electrical conductivity changes when the moon blocks the solar radiation.

The response in PBL due to the solar eclipse is much small compared to the responses of ionospheric layers since the ionization process in PBL is not directly related to the solar radiation and the electrical conductivity in the PBL much smaller than in the ionosphere, in which the relaxation time is $\sim 10^2$ sec nearby the surface but $\sim 4$ sec at 18 km and $\sim 10^{-4}$ sec at 70 km [13].

When the moon blocks the solar radiation, the electrical conductivity is getting reduced in the atmosphere. Since the ionization process in upper atmosphere is much more dependent to the solar radiation, the reduction of conductivity in the upper atmosphere is much bigger than in low atmosphere. Therefore, the net excess charge density in the low boundary of ionosphere should be lower in the lunar shadow (umbra and penumbra of the solar eclipse) than the outside of the shadow. These variations in ionosphere and PBL, in turn, affects to the telluric charge distribution to minimize the electric effect in the low atmosphere as shown in Fig. (8-b), in which the relaxation time is $\sim 10^{-5}$ sec [13].

If the solar PVN effect is uniform and constant during the solar eclipse, these charge distributions as shown in Fig. (8-b) should be followed along with the moon's shadow that sweeps across the earth's surface with velocity $\upsilon \sim 450 \text{ m/sec}$ from West to East. Since the electrical equilibrium state as in the Fig. (8-b) cannot be reached instantly but should be delayed due to the sweeping speed, there can be downward currents from the upper ionosphere and inward currents into the shadow region of the low boundary of ionosphere. However, the major contribution of these induced currents should be alongside the earth's magnetic field lines since the parallel conductivity is much higher than the Pedersen conductivity and Hall conductivity in ionosphere.

If the electrical conductivity underground for telluric charges is not affected by the solar eclipse but uniform at least in horizontal direction, the induced telluric charge distribution cannot be the same as in the low boundary of ionosphere, especially, nearby the boundary of penumbra (the first and the fourth contacts) in which the telluric charge density is smaller than the ionospheric charge density at the low boundary as shown in the Fig. (8-b).

As shown in section (3.1), gravitational field is not changed as long as the interaction strength to the induced positive charge distribution is the same but in the opposite direction as to the induced negative charge distribution.

Due to the charge density difference nearby the boundary of penumbra in the Fig. (8-b), net positive charge effect (at PBL including telluric charges) on the gravitational field is bigger than the negative charge effect (at the low boundary of the ionosphere); thus, it is appeared as the gravitational reduction on the surface of the earth.

If the electric field intensity nearby the low boundary of the ionosphere is $E \sim 10^{-6} \text{ V/m}$ [15], the electric field effect from the charge density difference nearby the boundary of penumbra can be estimated as $\sim 10^{-6} \text{-} 10^{-7} \text{ (V/m)}$. It is corresponded to $\Delta g \sim 10^{-7} \text{-} 10^{-8} \text{ (m/sec}^2\text{)}$ from the Eqn (6).

Actually, the gravity reduction in east side of penumbra area can be bigger than the west side area because of the earth's rotation. It has been observed that there is a slightly difference in the variation of the gravity in west side and east side of penumbra areas [24][26]. Furthermore, the variation of apparent vertical direction of the gravity field can be appeared nearby the edge of the penumbra [24][27]. All these variations of gravity should be dependent on the ionospheric condition, and the variations of electric field and magnetic field also should be accompanied.

## 6. 2 Plate tectonics and earthquakes

It has been told for a long time that geomagnetic storms are related to earthquakes [30], but still the scientific relation (if any) between these two distinct natural phenomena has not been cleared yet.

It has been known that the earth's geomagnetic field is closely affected by the space weather, especially, by solar activities [22]. When the solar activity increase, in which the number of sunspots is increased, a huge number of charged particles, for instance, from coronal mass ejections(CME) in the Sun come into the earth's magnetosphere, which arises the magnetic storms and affects the earth's internal magnetic field. Furthermore, the charge distribution in the ionosphere of the earth is changed, and then it affects the telluric charge distribution of the earth to minimize the gravity variation.

The hard crust of the earth keeps being pushed out by the magnetic pressure and, of course, by thermal pressure from inside the earth and pulled down by the gravitational force. However, the magnetic field inside the earth cannot be constant because the induced charges inside the earth are not isolated from outside but keep being fluctuated lively responding to the outside electric environment since a global electric circuit can be imagined as being connected from the earth's surface to ionosphere, and to polar areas. Hence, if the magnetic pressure is changed or become unstable, the force balance between the magnetic pressure and the gravitational force becomes unstable, and it can be one of major reasons for Plate tectonics of the earth [31].

Furthermore, the correlations of earthquakes to the variations of sunspots and variations in ionosphere [22] are also clear: If the magnetic storm hits the earth's magnetosphere like shock waves, it makes instability of the magnetic field inside the earth and the telluric charge distribution of the earth. This acute local unbalance across the crust between gravity and magnetic field inside the earth causes magma to move, which should be a reason for the earthquakes and volcanic eruptions besides the thermal activities inside the earth.

## 7. **Conclusions**

Gravitation and electromagnetism is not independent to each other; gravitational field and electromagnetic field has a unique ontological entity in 4-D complex space. On this respect, basic physical laws are reviewed, and the PVN (primitive-virtual negative) charge was introduced for the mass-charge interaction.

Since the strength of mass-charge interaction is so small, the interaction effect was searched in macroscopic phenomena. Firstly, the origin of electric field on the surface of the earth was explained naturally. Secondly, the origin of the earth's main magnetic field was explained with an alternate model. The model was also applied to other planets in solar system, in which the estimation with the model was compared with known data.

Moreover, it is expected from the model: Gravity in the earth's mantle is practically ignored, but magnetic pressure is dominant. With thermal pressure inside the earth, the magnetic pressure and the gravitational pressure on the earth's crust makes a mechanical force balance. Thus, the variation of earth's magnetic field should be one of reasons for the plate tectonics and the earthquakes, in which the variation of magnetic field is considerably affected by solar activities.

Not quantitatively but qualitatively and consistently, unsettled phenomena, such as antigravity phenomenon, gravitational anomalies during solar eclipses were explained.

## 8. Discussions

Since gravitational field is interacting with electric charges in 4-D complex space, the subject of dark matter [32], which has been one of puzzles in astronomy for a long time, is needed to review from the beginning. Moreover, the validation limit of equivalence principles in general theory of relativity is needed to be examined.

Considering that the transversal mode is for electromagnetic wave propagation and the longitudinal mode is for gravitational wave propagation in 4-D complex space, the speed of the longitudinal wave can be different from the speed of electromagnetic wave. However, an experimental conformation about the existence of the longitudinal wave should be necessary, first of all.

For Pioneer anomaly and flyby anomaly [33], the possible explanation is also expected when the PVN charge of the spaceflight is considered with the mass-charge interaction because there should be a net negative charge distribution at the edge of solar system (Pioneer anomaly) and there should be some difference in the ionospheric charge distribution at dayside and night side of the earth (flyby anomaly). However, the interaction of the spaceflight's PVN charge with the earth's magnetic field is too small to explain the flyby anomaly.

It is desirable to review natural phenomena with the concept of mass-charge interaction, especially natural disasters on the earth -- earthquakes, volcano, hurricane, tornado, etc. For example, once we can understand clearly the mechanism of tornado, we can think more progressively how to minimize tornado damage and how to suppress the tornado outbreak itself.

# References

[1] Kiyoung Kim, arXiv:quant-ph/9701023, e-print archive.

[2] David Bohm, *Physical Review*, vol. **85**, (1959), pp. 166; Goldstein, Sheldon, "Bohmian Mechanics", *The Stanford Encyclopedia of Philosophy (Fall 2008 Edition)*, Edward N. Zalta (ed.), URL = <http://plato.stanford.edu/archives/fall2008/entries/qm-bohm/>.

[3] Kiyoung Kim, arXiv:quant-ph/9706063, e-print archive.

[4] Donald H. Perkins, *Introduction to High Energy Physics*, (Addison-Wesley Publishing Company, Inc., 1987), pp. 23.

[5] Harald A. Enge, *Introduction to Nuclear Physics*, (Addison-Wesley Publishing Company, Inc., 1987), pp. 440; Arthur Beiser, *Perspectives of Modern physics*, (McGRAW-HILL KOGAKUSHA, LTD), pp. 502-503; Binding energy. (2008, October 18). In *Wikipedia, The Free Encyclopedia*. Retrieved 21:50, October 26, 2008, from <http://en.wikipedia.org/w/index.php?title=Binding_energy&oldid=246065405>.

[6] Eur. Phys. J. C 15, 1-878 (2000), pp. 53-54.

[7] Paul Lorrain & Dale R. Corson, *Electromagnetic Fields and waves*, (W.H. Freeman and Company, $2^{nd}$ edition, 1970) pp. 444.

[8] Saxl, E.J., "An Electrically Charged Torque Pendulum" *Nature*, July 11, 1964, p. 136; Thomas F. Valone, Progress in Electrogravitics and Electrokinetics for Aviation and Space Travel, Presented at the Space Tech. App. Info. Forum, Albuquerque, NM, Feb. 2006 ; more elsewhere.

[9] Bahram Mashhoon, arXiv:gr-qc/0311030v1, e-print archive.

[10] Anti-gravity. (2008, October 22). In *Wikipedia, The Free Encyclopedia*. Retrieved 21:56,October 26, 2008, from <http://en.wikipedia.org/w/index.php?title=Anti-gravity&oldid=246875380>.

[11] Kiyoung Kim, arXiv:hep-ph/9811522, e-print archive.

[12] *The Earth's Electrical Environment*, (National Academy Press, 1986), pp149; R. Latha, *Earth Planets Space*, **55**, 677-685, 2003.

[13] *The Earth's Electrical Environment*, (National Academy Press, 1986), pp 211, pp 214.

[14] *The Earth's Electrical Environment*, (National Academy Press, 1986), pp 232.

[15] *The Earth's Electrical Environment*, (National Academy Press, 1986), pp 191.

[16] Larmor, J., *Rep. Brit. Assoc. Adv. Sci.* 1919, **159**; Elsasser, W. M., *Physical Review*, 1939. **55**, p. 489-498; Herndon, J. M., *Current Science*, 2007, **93**(11), p. 1485-1487.

[17] The Interior of the Earth: Eugene C. Robertson. Last modified 05-21-07. <http://pubs.usgs.gov/gip/interior/>; Textbooks in Geophysics.

[18] Table of the Planets and Sun: Dr. Michael Collier, Last modified 12-12-203. <http://lepmfi.gsfc.nasa.gov/mfi/lepedu/planets.htm>.

[19] Zuber, Maria T. "Mercury." World Book Online Reference Center. 2004. World Book, Inc. (http://www.worldbookonline.com/wb/Article?id=ar356240.).

[20] Metallic hydrogen. (2008, October 24). In *Wikipedia, The Free Encyclopedia*. Retrieved 14:08, October 27, 2008, from http://en.wikipedia.org/w/index.php?title=Metallic_hydrogen&oldid=247362387 .

[21] Fraser-Smith, A.C. (2008), Eos Trans. AGU, 89(23); M. Kamogawa, Eos, Vol. 87, No. 40, pp. 417-424, (Oct. 2006); Masashi Kamogawa, *et al*., TAO, Vol. 15, No. 3, 397-412 (Sept., 2004); Naoto Ujihara, *et al*., *Earth Planets Space*, **56**, 115-123, 2004; Kazuhiko Teramoto and Motoji Ikeya, Jpn. J. Appl. Phys. **39** (2000) pp. 2876-2881; T. Sengor, *Proc. of PIERS*, Cambridge, USA, 904 (July 5-14, 2000); more elsewhere.

[22] Solar Activity: Dr. Tycho von Rosenvinge. Last modified: 09-09-2008. <http://helios.gsfc.nasa.gov/solaract.html> .

[23] J. H. Gundlach and S. M. Merkowitz, Phys. Rev. Lett. **85** 2869 (2000); J. B. Fixler, *et al*., Science 5 January 2007: Vol. 315. no. 5808, pp. 74-77; Riley Newman, *Convergence (?) of G Measurements – Mysteries Remain*., Retrived: 10-27-2008. < http://www.phys.lsu.edu/mog/mog21/node12.html> .

[24] Chris P. Duif, arXiv:gr-qc/0408023, e-print archive; Xavier E. Amador, *Journal of Physics: Conference Series* **24** (2005) 247–252.

[25] D.C. Mishra and M.B.S. Rao, Curr. Sci. **72**, 782 (1997)

[26] Q. -S. Wang, X.-S.Yang, C.-Z. Wu, H.-G. Guo, H.-C. Liu and C.-C. Hua, Phys. Rev. D **62** (2000) 041101.

[27] T. Kuusela, *et al*., Phys. Rev. D **74,** 122004 (2006)

[28] H. Chandra, *et al.*, Astronomical Society of India, BASI, Vol. 8, No. 4, pp. 149-151; R.Sridharan, *et al*., Annales Geophysicae (2002) 20: 1977-1985; H. Chandra, *et al*., *Earth Planets Space,* Vol. 59 (No. 1), pp. 59-64, 2007; more elsewhere.

[29] Lilley, F.E.M. and Woods, D.V., Journal of Atmospheric and Terrestrial Physics, Vol. 40, pp. 749-754; A. Adam, J. Vero, and J Szendroi, Solar eclipse effect on geomagnetic induction parameters, Annales Geophysicae, 23, 3487-3494, 2005; more elsewhere.

[30] Motoji Ikeya, *Earthquakes and Animals: From Folk Legends to Science*, (World Scientific Publishing Co. Pte Ltd., 2004).

[31] Pidwirny, Michael (Lead Author); John Shroder and Galal Hassan Galal Hussein (Topic Editors). 2007. "Plate tectonics." In: Encyclopedia of Earth. Eds. Cutler J. Cleveland (Washington, D.C.: Environmental Information Coalition, National Council for Science and the Environment). [Published in the Encyclopedia of Earth January 30, 2007; Retrieved December 28, 2007]. <http://www.eoearth.org/article/Plate_tectonics>

[32] D. W. Sciama, *Modern Cosmology and the Dark Matter Problem*, (Cambridge University Press 1993), part one.

[33] John D. Anderson, James K. Campbell, Michael Martin Nieto, New Astronomy, Vol. 12, Issue 5, July 2007, pp. 383-397;
Pioneer anomaly. (2008, October 24). In *Wikipedia, The Free Encyclopedia*. Retrieved 14:05, October 27, 2008, from
http://en.wikipedia.org/w/index.php?title=Pioneer_anomaly&oldid=247309455